\documentclass[12pt]{iopart}
\usepackage{iopams} 
\usepackage{mathptmx}
\usepackage[utf8]{inputenc} 
\usepackage{graphicx}
\usepackage{epstopdf}
\usepackage[numbers,comma,square,sort&compress,merge]{natbib} 
\usepackage{textcomp}
\usepackage{xspace}
\usepackage[innercaption,ragged]{sidecap}
\usepackage{url}
\usepackage{eurosym}
\usepackage{color,soul} 
\usepackage[colorinlistoftodos,prependcaption,textsize=footnotesize,textwidth=1.4cm]{todonotes}

\usepackage{hyperref}
\hypersetup{pdftitle={Growth mechanisms in molecular beam epitaxy for GaN-(In,Ga)N core-shell nanowires emitting in the green spectral range}, colorlinks, citecolor=blue}

\newcommand{\grad}{$^{\circ}$\xspace}
\newcommand{\celsius}{$^{\circ}$C\xspace}

\begin{document}

\title[Growth mechanisms for GaN-(In,Ga)N core-shell nanowires]{Growth mechanisms in molecular beam epitaxy for GaN-(In,Ga)N core-shell nanowires emitting in the green spectral range}
\author{David van Treeck\footnote{These authors contributed equally to the work.}, Jonas L\"ahnemann\footnotemark[\ddagger], Oliver Brandt, Lutz Geelhaar}
\ead{laehnemann@pdi-berlin.de}
\address{Paul-Drude-Institut für Festkörperelektronik, Leibniz-Institut im Forschungsverbund Berlin e.V., Hausvogteiplatz 5--7, 10117 Berlin, Germany}

\begin{abstract}
Using molecular beam epitaxy, we demonstrate the growth of (In,Ga)N shells emitting in the green spectral range around very thin (35~nm diameter) GaN core nanowires. These GaN nanowires are obtained by self-assembled growth on TiN. We present a qualitative shell growth model accounting for both the three-dimensional nature of the nanostructures as well as the directionality of the atomic fluxes. This model allows us, on the one hand, to optimise the conditions for high and homogeneous In incorporation and, on the other hand, to explain the influence of changes in the growth conditions on the sample morphology and In content. Specifically, the impact of the V/III and In/Ga flux ratios, the rotation speed and the rotation direction are investigated. Notably, with In acting as surfactant, the ternary (In,Ga)N shells are much more homogeneous in thickness along the NW length than their binary GaN counterparts.
\end{abstract}



\submitto{\NT}

\ioptwocol

\maketitle

\section{Introduction}

Devices based on three-dimensional (3D) nano- and microstructures with radial core-shell designs for the active region have been demonstrated for a number of applications \cite{Mata2013,Mandl2013,Dasgupta2014}.
One major advantage of core-shell nanowires (NWs) over planar structures is that the ratio of active region area per footprint area can be dramatically increased by increasing their aspect ratio. This makes core-shell structures particularly interesting for applications such as solar cells or light emitting diodes (LEDs), where large active regions are advantageous \cite{Waag2011,LaPierre2013a}.
In particular with the increasing demand for compact high-resolution, multicolour displays, LEDs based on nano- and microstructures will play an important role in the next generation of display technologies \cite{Barrigon2019,Wong2019}.

For \textmu-rod-based GaN-(In,Ga)N core-shell LEDs with much larger diameters than NWs, the fabrication processes have been successively optimised over the last years \cite{Koester2015,Schimpke2016,Kapoor2018}. Although the growth of blue core-shell \textmu-rod LEDs by metal-organic chemical vapour deposition (MOCVD) is well established \cite{Koester2015,Schimpke2016} and emission up to the red spectral region has already been demonstrated for axial (In,Ga)N/GaN multi quantum wells (MQWs) grown by molecular beam epitaxy (MBE) \cite{Kishino_procspie_2007,Guo2010,Bavencove2010,Armitage2010,Nguyen2011b,Limbach2012a}, the growth of (In,Ga)N-based core-shell structures with high In contents emitting in the green to red colour spectrum remains a major challenge. In general, fundamental aspects such as the different nature of In and Ga regarding atom size and bond strength to N, as well as the high vapour pressure of N over InN, inhibit high In contents in (In,Ga)N/GaN heterostructures \cite{Ambacher1998a,Adelmann1999,Siekacz2011,Duff2014}. Especially for MOCVD growth, the incorporation of high In contents in GaN-(In,Ga)N core-shell structures seems to be far from being straightforward due to the usually lower In incorporation efficiency on the \textit{M}-plane sidewalls compared to the \textit{C}-plane \cite{Neubert2005a,Wernicke2012k,Kapoor2018}.

Kapoor \emph{et al.} \cite{Kapoor2018} demonstrated the growth of a green GaN-(In,Ga)N \textmu-rod core-shell LED grown by MOCVD emitting at 550~nm. MBE is often said to have a great potential for achieving shells with high In content on \textmu-rods or NWs. However, so far only violet emission has been reported for MBE-grown (In,Ga)N shells on \textmu-rod core structures obtained by MOCVD \cite{Albert2014, Albert2015a}. In order to push the In content to higher levels and achieve emission at longer wavelengths, a more comprehensive and fundamental knowledge of the peculiarities of (In,Ga)N shell growth by MBE is required.

We have recently discussed in detail the growth dynamics of binary GaN shells on GaN NW cores in relation to the directionality of deposition, substrate rotation, and the geometrical arrangement of the material sources in an MBE system \cite{vanTreeck2020}. In this contribution, we extend our work to the growth of ternary (In,Ga)N shells. To this end, we make use of the self-assembled GaN NWs on TiN developed in Ref.~\citenum{vanTreeck2018} as template NW cores. In contrast to other substrates, the TiN film facilitates long and well separated NWs as prerequisite for shell growth \cite{vanTreeck2020}. Compared with MOCVD-grown \textmu-rods, these NWs have an average diameter of only 35 nm. Based on the findings of Ref.~\citenum{vanTreeck2020}, we develop a qualitative model for the (In,Ga)N shell growth by MBE, which enables us to find growth conditions for achieving homogeneous shells emitting in the blue/green spectral range. Furthermore, we analyse the influence of the V/III and In/Ga ratio, as well as the impact of substrate rotation on the NW morphology and the emission properties of the (In,Ga)N shells.

\section{Methods}

\begin{figure}
\centering
\includegraphics[width=0.66\columnwidth, trim= 0 0 0 0]{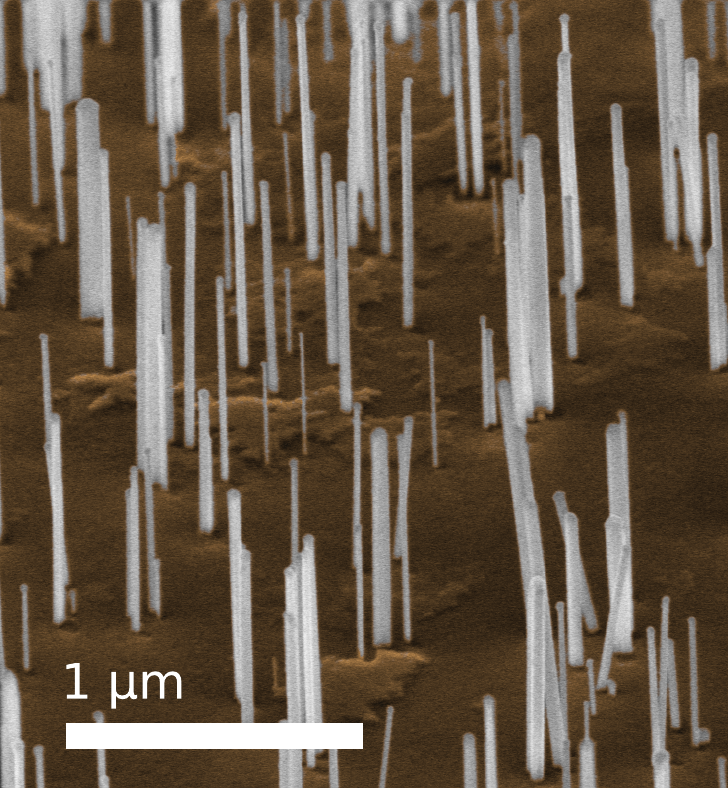}
\caption{Bird's eye SE micrograph of a representative self-assembled template NW ensemble prior to shell growth. The TiN film on the substrate is colorized for illustrative purposes.}
\label{fig1}
\end{figure}

The self-assembled GaN core NWs used as templates for the (In,Ga)N shell growth were synthesised by plasma-assisted MBE at a V/III ratio of 2.5 and a substrate temperature of 780~\celsius on nitridised Ti films on Al$_2$O$_3$ (0001) substrates. Substrate temperatures were determined with a calibrated pyrometer as described in Ref.~\cite{vanTreeck2020}. Further details on the core growth can be found in Refs.~\citenum{vanTreeck2018,vanTreeck2020}. The resulting GaN NW ensembles have a density of about 10$^9$~cm$^{-2}$ and a mean diameter and length of about 35~nm and 790~nm, respectively. The NW axis is oriented along the $[000\bar{1}]$ direction, and the six sidewalls are formed by \emph{M}-plane facets. A representative scanning electron (SE) micrograph of such a template NW ensemble is displayed in figure~\ref{fig1}. The variation in height and diameter of individual NWs is a consequence of the long incubation time for self-assembled growth on TiN, where their growth times vary as their nucleation is spread over a longer time period.

The shells of all samples presented in this study were grown for 50~min at a substrate temperature of 490\,\celsius. This temperature was established in a preliminary experiment, in which the substrate temperature was gradually lowered starting from 600\,\celsius and a monolayer (ML) of In was deposited and subsequently desorbed every 20\,\celsius until a temperature was reached at which no In desorption was visible anymore by line-of-sight quadruple mass spectrometry (QMS). The absence of In desorption ensured that all the In deposited on the sample was incorporated. 
In general, due to the low dissociation temperature of InN, only sufficiently low substrate temperatures allow high In contents \cite{Grandjean1998, Averbeck1999}. Moreover, high and homogeneous In contents may also be prevented by strain-related issues resulting from the different atomic size of Ga and In \cite{Duff2014}. For sample A, the growth parameters for the (In,Ga)N shells were a V/III ratio of 5  and an In/Ga flux ratio of 1 at a rotation speed of 10~rpm. For additional samples, individual parameters were varied: for sample B a V/III ratio of 2.5, for sample C an In/Ga flux ratio of 3 and for sample D a rotation speed of 1~rpm were chosen, while keeping all other parameters similar to sample A. The total provided metal flux supplied was always kept at 0.2~ML/s

The NW morphology was investigated in a Hitachi S4800 field-emission scanning electron microscope (SEM). Photoluminescence (PL) spectra at 300~K were acquired in top-view geometry on as-grown NW ensembles using a Horiba Jobin Yvon LabRAM HR 800 UV confocal \textmu-PL setup with excitation from a HeCd laser (325~nm). The spot diameter of about 3~\textmu{}m using a $15\times$ objective for the measurements at 300~K corresponds to the excitation of about 70 NWs. The PL was directed into a 0.8~m focal length monochromator equipped with a 600~lines/mm grating for spectral dispersion, and detected with a liquid-N-cooled charge-coupled device (CCD) camera. The emission along the axis of single, dispersed NWs was measured by cathodoluminescence (CL) spectral imaging using a Gatan MonoCL4 system mounted on a Zeiss Ultra55 field-emission SEM equipped with a cold-stage for measurements at 10~K. The low sample temperature facilitated faster scanning times during hyperspectral mapping. Individual spectra were recorded while scanning the electron beam with a step size of 20~nm to form a line scan. An acceleration voltage of 5~kV and a probe current of 1--1.5~nA were chosen. The light was collected by a parabolic mirror mounted above the sample and directed into a 0.3~m focal length monochromator with an entrance slit width of 1~mm using a 300~lines/mm grating. For detection, a liquid-N-cooled CCD was used, integrating each spectrum for 1~s. For CL measurements, the NWs were dispersed on a Au-coated Si substrate. Representative line scans for each sample were selected based on the room-temperature ensemble PL spectra. The CL data was analyzed and visualized using the python packages \texttt{HyperSpy} and \texttt{LumiSpy} \cite{HyperSpy,LumiSpy}.

\section{Results}
\subsection{Growth model for ternary (In,Ga)N shells}

\begin{figure}
\centering
\includegraphics[width=0.85\columnwidth, trim= 10 0 10 0]{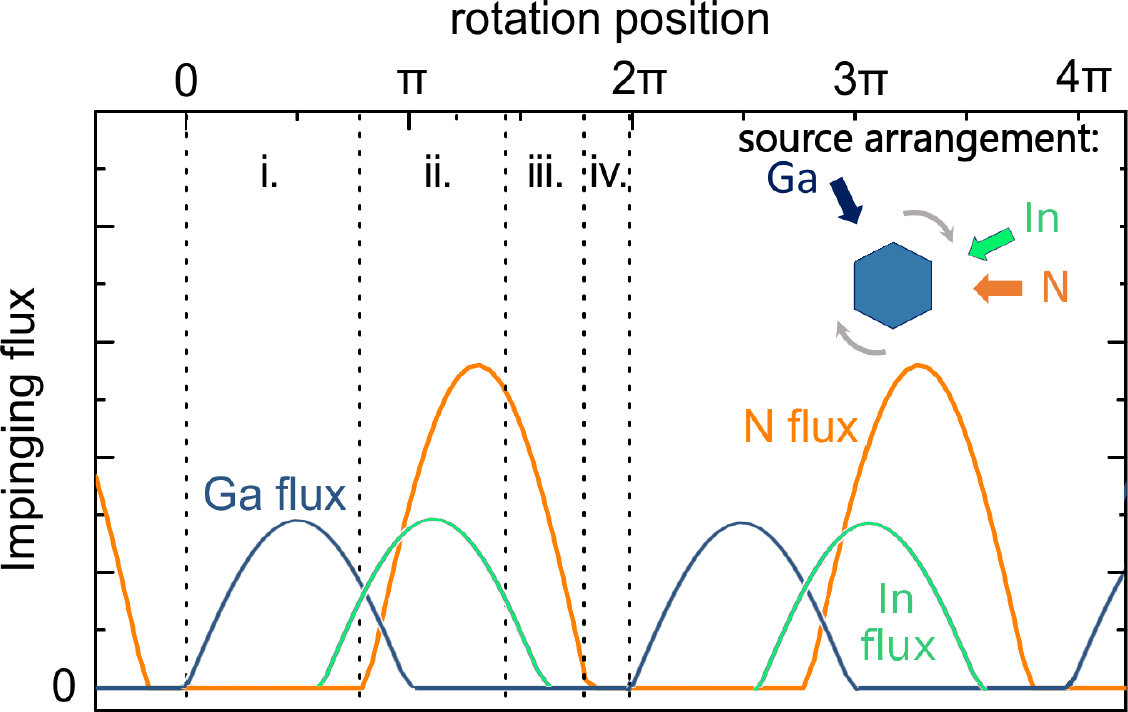}
\caption{Qualitative diagram of the In, Ga and N fluxes impinging on the side facet of a NW at different rotational positions (and thus different times during the rotation cycle) according to a geometrical model, assuming V/III and In/Ga flux ratios of 1. This graph illustrates the general time dependence of the different fluxes with respect to each other. The inset depicts the geometrical arrangement of the material sources in our MBE system, where the azimuthal angles between the different sources are $\beta_\mathrm{Ga-N}=144$\grad, $\beta_\mathrm{In-N}=36$\grad and $\beta_\mathrm{Ga-In}=108$\grad. The direction of rotation is indicated by the grey arrows.}
\label{fig2}
\end{figure}

To begin with, we propose a qualitative model for the growth of (In,Ga)N shells by MBE based on the insights gained for GaN shells in Ref.~\citenum{vanTreeck2020}. This model is based on the fact that the different molecular beams impinge on the NW side facets in an alternating sequence as the substrate is rotated. In contrast, the top facet and the substrate (if not shadowed) are constantly exposed to all species. The resulting difference in chemical potentials on the different facets induces adatom diffusion and variations in the adatom concentrations of the elements involved.

For the case of (In,Ga)N, figure~\ref{fig2} shows a qualitative diagram of the Ga, In and N fluxes impinging on a particular side facet with proceeding growth time during two rotation cycles. This sequence results from the substrate rotation for the cell arrangement of our MBE system as shown in the inset. As for GaN, we assume that the diffusion of the adatoms between adjacent side facets is negligible compared to the strong diffusion towards the NW top and the substrate \cite{Lymperakis2009}. As the following discussion will show, this hypothesis gives a reasonable description of (In,Ga)N shell growth.

As sketched in figure~\ref{fig2}, during each rotation of the (In,Ga)N deposition, a side facet passes through the four different phases established in Ref.~\citenum{vanTreeck2020}. The round starts with the \textit{wetting phase} (i.), where Ga is deposited on a side facet shadowed from both In and N. As the facet rotates away from the Ga source, Ga adatoms diffuse towards the NW top and the substrate, which are constantly exposed to N, resulting in gradients in the chemical potential. Once the side facet rotates into the In beam and In is deposited, we expect the diffusivity of the Ga adatoms to be dramatically enhanced due to In acting as surfactant \cite{Widmann1998a,Neugebauer2003a,Landre_apl_2008,Woelz_apl_2011}. As a consequence, a rather homogeneous distribution of Ga and In along the entire NW length can be assumed, even in the case that the lower part of some NWs is shadowed from the direct flux by adjacent NWs.

Second comes the \textit{metal-rich growth phase} (ii.), where the side facet rotates into the N beam and growth starts, but the V/III ratio on the facet remains below one. Only with further rotation into the N beam, the \textit{N-rich growth phase} (iii.) is reached. It should be noted that during the metal-rich growth phase, a lower In incorporation can be expected compared to the N-rich growth phase. Due to the different bond strengths of Ga and In to N, Ga is preferentially incorporated over In. Thus, under metal-rich conditions, the maximum In incorporation is determined by the difference in N and Ga supply \cite{Averbeck1999,Duff2014}. Furthermore, it has been found that under N-rich conditions, the In incorporation increases further with higher V/III ratios \cite{Siekacz2011,Lang2012a}.

The final phase is the \textit{dissociation phase} (iv.). Depending on the arrangement of the sources, there may be an additional rotation phase during which the side facet is not exposed to any of the atomic fluxes. Due to the high vapour pressure of N over InN \cite{Morkoc2008a}, the previously grown (In,Ga)N on this facet may thermally dissociate.

This model gives us a guideline on the growth conditions required to achieve a high In incorporation. Since the In incorporation is expected to be reduced during the metal-rich growth phase due to the diffusion of In away from the side facet, keeping this phase short should allow for higher In contents. To this end, we aim for a high V/III ratio and a high rotation speed. The latter has the added benefit of shortening the dissociation phase. Moreover, to obtain a high In content, the shell growth should be carried out at a substrate temperature where In desorption and dissociation are kept to a minimum.

\subsection{Towards high In content (In,Ga)N shells}

\begin{figure}
\includegraphics[width=1\columnwidth, trim= 0 0 0 0]{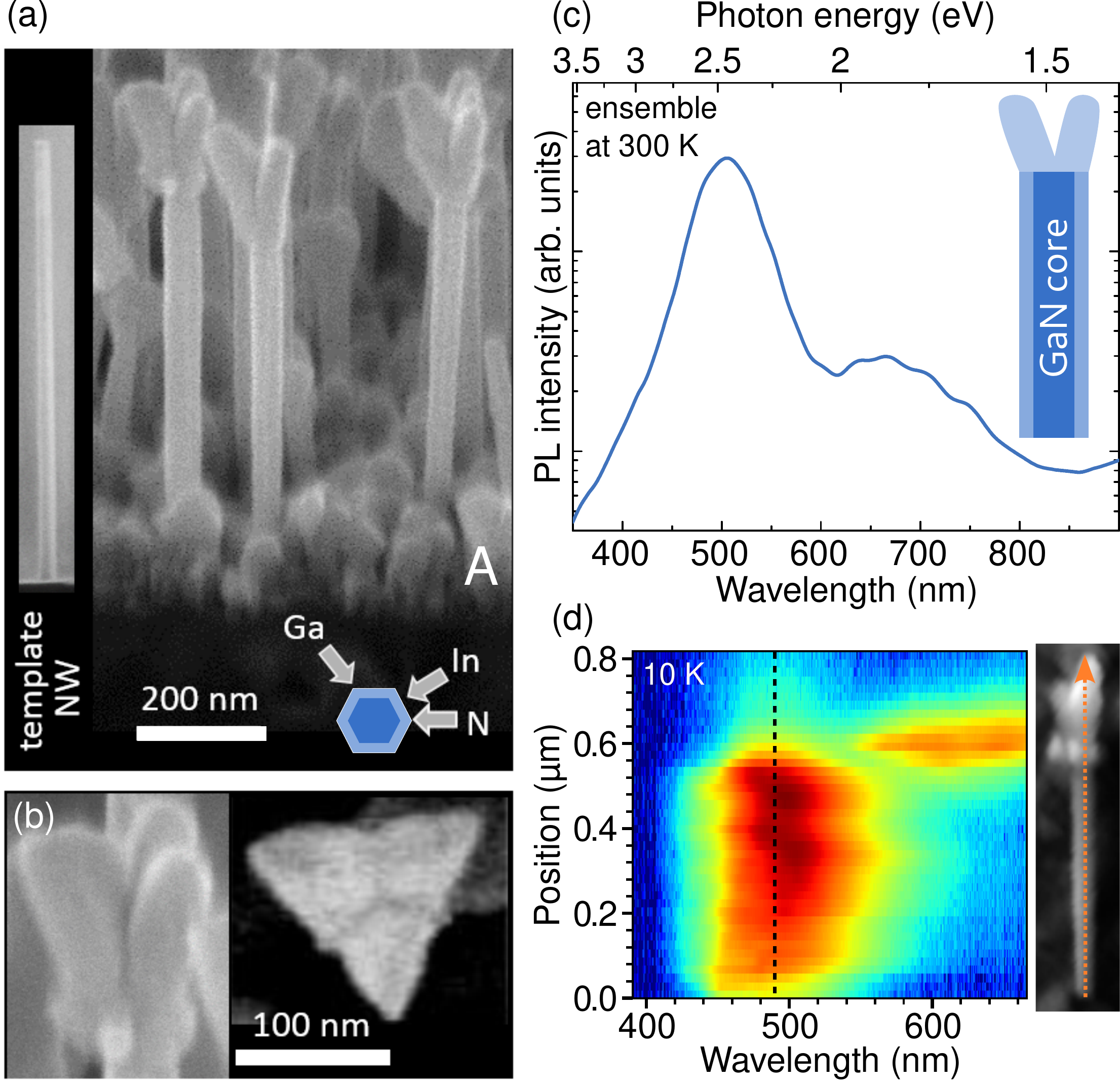}
\caption{(a) Bird's eye SE micrograph of sample~A in which an (In,Ga)N shell was grown around GaN NW cores (\emph{cf.} left inset). The lower right inset sketches the source arrangement. (b) Bird's eye and plan view SE micrographs of the broad top segments found for the NWs of sample~A. (c) Ensemble PL measured at 300~K. The inset sketches the core shell structure with higher In content indicated by lighter blue tones. (d) CL line scan of a single exemplary NW measured at 10~K alongside the corresponding SE micrograph with the scan position indicated by an orange arrow. Note that all line scans in this manuscript are plotted on the same logarithmic colour scale.}
\label{fig3}
\end{figure}

The practical implementation of these considerations for the growth of (In,Ga)N shells with a high In content is shown in figure~\ref{fig3}. The bird's eye SE micrograph in figure~\ref{fig3}(a) shows the morphology of sample~A containing GaN NWs with (In,Ga)N shells of about 10--15~nm thickness---evaluated from SE micrographs for an average diameter of the GaN NW cores of 35~nm. The substrate rotation speed was 10~rpm.\footnote{Limited by the driving motor for the substrate rotation.} The V/III and the In/Ga flux ratios were 5 and 1, respectively, with $\Phi_\mathrm{In}=\Phi_\mathrm{Ga}=0.1$~ML/s. For our MBE system, this V/III ratio is about the highest possible value while maintaining a reasonably high In and Ga flux. The resulting (In,Ga)N shells are fairly homogeneous in thickness along the entire length of the NWs. However, large tripod-shaped top segments have formed, as visible in the magnified SE micrographs in figure~\ref{fig3}(b). Additionally, parasitic (In,Ga)N growth is visible on the substrate between the NWs.

The homogeneity of the (In,Ga)N shell thickness, in contrast to the hourglass-shaped GaN shells in Ref.~\citenum{vanTreeck2020} and despite the low substrate temperature as well as the nominally very N-rich growth conditions, may be explained by In acting as a surfactant and enhancing the diffusivity of Ga adatoms on the \textit{M}-plane side facets \cite{Widmann1998a,Landre_apl_2008,Woelz_apl_2011}. The effect might be similar to the one found for the (0001) facet, where In forms an adlayer on the surface. This adlayer enhances the mobility of Ga adatoms and leads to smooth layers even at low substrate temperatures \cite{chen2001, Neugebauer2003a}. In the presence of In, the diffusion of metal adatoms around the NW, from one facet to the next, might also be increased leading to a more homogeneous adatom concentration along the NW.

The formation of the pronounced top segments may be explained by the increased diffusivity and strong gradient in chemical potential due to the highly N-rich growth conditions, as all fluxes impinge simultaneously on the top facet. This situation leads to strong adatom diffusion towards the NW top and results in a high growth rate on the top facet. The tripod shape of these top segments may be ascribed to cubic (In,Ga)N growing along the three (111) axes of the zincblende structure \cite{Zhang2016u}. In general, low substrate temperatures ($<600$\,\celsius) favour the formation of the cubic polytype as well as of stacking faults in nitride materials \cite{Renard2010,Jacopin2011}. Moreover, the addition of In has been found to promote the formation of the cubic polytype \cite{Kim2005}.

Figure~\ref{fig3}(c) shows a room-temperature PL measurement of the NW ensemble of sample~A. The spectrum shows an intense emission band at about 505~nm (2.45~eV) with a full width at half maximum (FWHM) of about 68~nm (340~meV) and a weaker emission band at about 665~nm (1.86~eV) with an even broader FWHM. In order to elucidate the spatial origin of the different emission bands, CL line scans along single, dispersed NWs were acquired at 10~K. Figure~\ref{fig3}(d) depicts a characteristic line scan showing a rather homogeneous emission at about 490~nm (2.53~eV) along the whole length of the NW. Remaining variations of the emission wavelength and band width are attributed to compositional fluctuations in the ternary alloy. The second emission band, which peaks above 600~nm, is only visible in the area of the enlarged top segment. The parasitic (In,Ga)N layer grown between the NWs also emits in the spectral region of the top segments, although its emission intensity is comparatively low (not shown). Note that due to the PL measurements being performed in top-view geometry on the NW ensemble, the contribution of the top-segments is clearly enhanced in these spectra. Overall, the CL line scan clearly shows that the dominant emission band can be attributed to the (In,Ga)N shells. 

With sample~A, we demonstrate that MBE-grown GaN-(In,Ga)N core-shell NWs with shells emitting in the blue/green spectral range at around 505~nm can be achieved. Using the emission energies of planar (In,Ga)N layers as a reference \cite{Schley2007}, the peak luminescence would correspond to an In content in the shell of around 25\%. Note that this estimate does not take into account the differences in strain state between a planar (In,Ga)N layer and the (In,Ga)N shells.

In general, the emission wavelength fluctuations between NWs observed by CL (see section~\ref{sec:diam}) are much smaller than those observed for MBE-grown axial heterostructures on self-assembled NW ensembles on Si, where the emission wavelength varies from blue to red among individual NWs \cite{Kikuchi_2006,Lahnemann2011a,Bavencove_nt_2011,Limbach2012a,vanTreeck2019}. The absence of any luminescence of the GaN core in the PL spectra indicates that the electron-hole pairs excited in the core can efficiently diffuse to the (In,Ga)N shell and the top segment. The rather broad emission band of the (In,Ga)N shell is attributed to local fluctuations of the In content, resulting in charge carrier localisation and recombination at different wavelengths. The absence of sharp emission peaks indicating localisation in the low-temperature PL and CL spectra of single NWs may be due to the high excitation density. Moreover, the (In,Ga)N shell shows intense luminescence even without passivation by an outer GaN shell. This observation is another indication of carrier localisation, which prevents the charge carriers from reaching the surface and recombining non-radiatively. A rather broad emission band was also observed for MBE-grown planar \textit{M}-plane (In,Ga)N/GaN QW structures \cite{Sun_prb_2003} and (In,Ga)N shells around GaN \textmu-rods \cite{Albert2014} emitting in the violet spectral range. In addition, atom probe tomography of MOCVD-grown GaN-(In,Ga)N core-shell structures emitting in the green spectral range revealed a strong variation in the In content of the \textit{M}-plane shells \cite{Coulon2017,Kapoor2018}.

The long-wavelength emission of the top segment can be attributed to a higher In incorporation compared to the shell. In general, the incorporation efficiency of In on the N-polar \textit{C}-plane as well as on the semi-polar \emph{R}-planes is expected to be higher than that on the \textit{M}-plane \cite{Neubert2005a,Nath2011,Wernicke2012k,Duff2014}. However, from the CL analysis it is not clear from which part of the top segment the high-wavelength emission originates, \emph{i.e.}\ whether the luminescence stems from wurtzite or cubic (In,Ga)N. A possible explanation for the lower emission intensity of the top segments with respect to the shell, despite their large volume, could be a higher density of non-radiative defects.

\subsection{Influence of V/III and In/Ga ratio}

\begin{figure}
\includegraphics[width=\columnwidth]{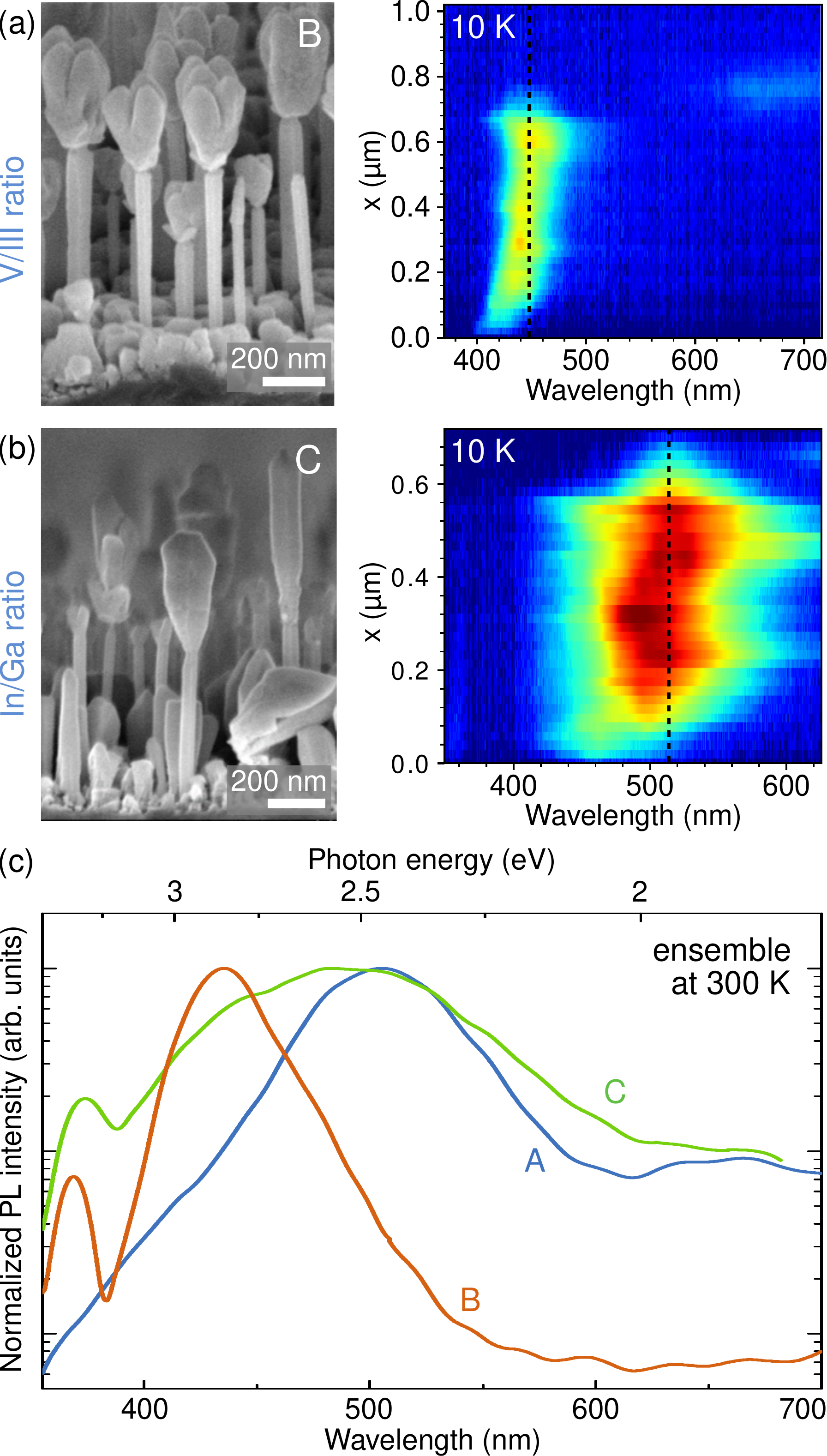}
\caption{Bird's eye SE micrograph and CL line scan measured at 10~K for (a) sample~B and (b) sample~C, which were grown with a V/III ratio of 2.5 and an In/Ga ratio of 3, respectively. (c) PL of the NW ensembles of sample~A, B, and C.}
\label{fig4}
\end{figure}

To investigate the influence of the V/III flux ratio, sample~B was grown at half the N flux $\Phi_\mathrm{N}$ and thus a V/III ratio of 2.5, while keeping all other growth parameters unchanged. As seen in figure~\ref{fig4}(a), this change results in a slight increase in the shell thickness to about 15--20~nm with a slight taper towards the substrate. The tripod-shaped top segment is even larger than for sample A. The CL line scan of a representative NW shows a weaker and less homogeneous shell emission, with a gradient from about 420~nm (2.95~eV) at the NW bottom to about 445~nm (2.79~eV) at the NW top, and a very weak top segment emission at about 670~nm (1.85~eV).

The shorter emission wavelength of the (In,Ga)N shell compared to sample~A points to a lower In incorporation on the NW \textit{M}-planes at lower V/III ratio. This effect has previously been observed both for planar MBE-grown (In,Ga)N films on \textit{C}-plane GaN as well as for MBE-grown (In,Ga)N shells on \textmu-rods \cite{Albert2014} and was attributed to a lower dissociation rate of (In,Ga)N with increasing N flux \cite{Siekacz2011,Lang2012a}.

Besides this direct influence of the V/III ratio on the stability of (In,Ga)N, the combination of the MBE geometry with the 3D nature of the nanostructures may further reduce the In incorporation on the NW side facets. In Ref.~\citenum{vanTreeck2020}, we have shown that a lower nominal V/III ratio in the growth of GaN shells also leads to a decreased metal incorporation on the side facets. As discussed above, once the side facet is rotated into the N beam, the growth conditions initially remain metal-rich. During this stage, since Ga is preferentially incorporated over In, the In incorporation on the NW sidewall may be rather low and excess In may diffuse towards the N-rich NW top facet and substrate. Such diffusion leads to an inhomogeneous In adatom concentration on the side facet, resulting in different shell thicknesses and In contents along the NW, once N-rich conditions are reached upon further rotation. Indeed, the shell is thicker at the top of the NW and the CL line scans show that the emission wavelength of the (In,Ga)N shell is red-shifted towards the top, suggesting an increased In content. In general, the lower the V/III ratio, the longer it takes to establish N-rich conditions favouring In incorporation and the lower the overall In content.

The In/Ga ratio is another important factor for the growth of (In,Ga)N. Sample C, shown in figure~\ref{fig4}(b), was grown with an In/Ga ratio of 3 keeping the total metal flux the same at 0.2 ML/s and otherwise similar growth conditions to those used for sample~A. The shell thickness is homogeneous along the NW length, but the approximately 5--10~nm thick shells are slightly thinner than those of sample~A. The shapes of the top segments are more diverse. About half of the NWs have enlarged top segments that grow at the rate of the incident N flux (N-limited conditions). The other half have only very small tripod-shaped top segments. The volume of the parasitic layer between the NWs has also increased significantly compared to the previous samples.

The CL line scan of a dispersed NW of sample~C (small top segment) shows the shell emission at about 515~nm (2.41~eV). In contrast to sample~A, we found a strong variation of the emission wavelength in the range of 420--525~nm (2.88--2.36~eV) between different NWs.

Figure~\ref{fig4}(c) shows the room-temperature PL spectra of samples~B and C in comparison with sample~A. With a peak emission wavelength of around 435~nm (2.85~eV) and a FWHM of 37~nm (250~meV), the main emission band for sample~B resembles well the (In,Ga)N shell emission found for CL line scans of single NWs at low temperature. Hence, the ensemble PL confirms the blue shift of the shell emission with decreasing V/III ratio. Consistent with the low intensity in the CL line scan, there is no additional PL band attributable to the top segments. The ensemble PL of sample~C exhibits a peak emission of the (In,Ga)N shells at about 485~nm (2.56~eV), slightly lower than that of sample~A. The broad FWHM of about 115~nm (640~meV) represents the variation observed between individual NWs.

The CL and PL analysis of sample C shows that the higher In/Ga ratio does not lead to an increase in the average In content of the (In,Ga)N shells, but rather to greater variations from NW to NW. Similarly, Albert \emph{et al.} \cite{Albert2014} observed no significant change in the In content of the (In,Ga)N shells around GaN \textmu-rods with increasing In/Ga ratio. This observation is in contrast to that for planar MBE-grown (In,Ga)N films on \textit{C}-plane GaN, where an increase in In content with increasing In/Ga ratio was found under similar N-rich conditions \cite{Bottcher1998,Storm2001}. This observation raises the question of whether the missing increase in In incorporation on the NW \textit{M}-planes is due to the 3D nature of the NWs (\textmu-rods). If the extra amount of In is not incorporated in the shell, it must be incorporated elsewhere as QMS shows no In desorption. For sample~C this could explain the large amount of material growing on the NW top facets and the increased parasitic growth on the substrate. The diffusion of the additional In away from the side facets may be attributed to a lower In incorporation efficiency on the \textit{M}-plane. The massive top segments can result in shadowing effects, which in turn may explain the strong variations in the emission of individual NWs and the diversity of the top segments.

\subsection{Influence of substrate rotation}

\begin{figure}
\includegraphics[width=\columnwidth, trim= 0 0 0 0]{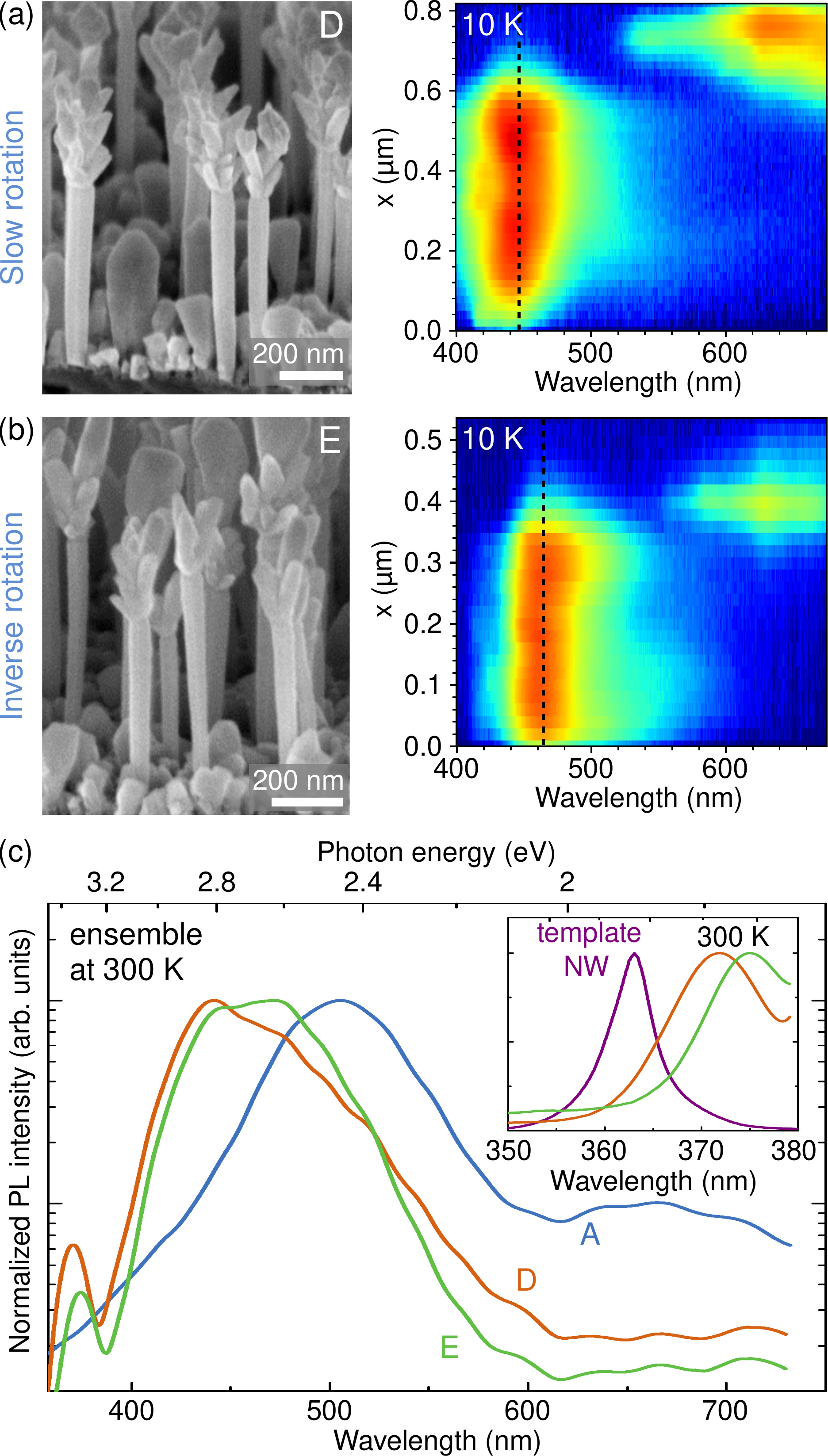}
\caption{Bird's eye SE micrograph and CL line scan measured at 10~K for (a) sample~D and (b) sample~E, which were grown with slow rotation (1~rpm) and inverse rotation (with respect to sample~A), respectively. (c) PL of the NW ensembles of sample~A, D, and E. The inset shows the GaN core PL of the template NW ensemble as well as of sample~D (orange) and E (green) at 300~K.}\label{fig5}
\end{figure}

In Ref.~\citenum{vanTreeck2020}, we found a strong influence of the rotation speed on the morphology of GaN shells. In the case of (In,Ga)N, the effect of changing the growth parameters on the homogeneity of the shell thickness seems to be less pronounced. However, the previous section has shown that the growth conditions can strongly influence the In incorporation.

Sample~D, shown in figure~\ref{fig5}(a), was grown at a substrate rotation speed ten times lower than that of sample~A, namely 1~rpm, at otherwise similar growth conditions. The shell morphology of the NWs is very similar to that of sample~A, while the top segments in most cases show a double tripod-shape and the parasitic growth is more pronounced. According to the CL line scan of a representative NW, the (In,Ga)N shell emits at about 445~nm (2.79~eV) with a FWHM of 67~nm (430~meV). The top segments show a rather broad emission at about 660~nm (1.88~eV). The lower emission wavelength of the (In,Ga)N shell compared to sample~A is confirmed by room-temperature PL measurements in figure~\ref{fig5}(c) and can be explained by a lower In incorporation at the slower substrate rotation.

Conceptually, lowering the rotation speed has a similar effect as lowering the V/III ratio. It takes longer until the metal-rich rotation phase is overcome and N-rich conditions are established on a particular side facet, increasing the chance for In to diffuse away and thus favouring Ga incorporation. The rotation speed can also affect the thermal dissociation of the (In,Ga)N shell when the facets are not exposed to any material beams.

For sample~E, shown in figure~\ref{fig5}(b), the direction of rotation was reversed in respect to sample~A (counter-clockwise). Again, the NWs look very similar to those of sample~A. However, the emission wavelength of the shell is at around 465~nm (2.67~eV) slightly lower, as revealed by the CL line scan and confirmed by the PL measurements in figure~\ref{fig5}(c). The FWHM of this emission band is 75~nm (420~meV). In this scenario, reading figure~\ref{fig2} from the right, In is partially co-deposited with N. However, as Ga is deposited after In, it is likely that the Ga adatoms replace the already bound In adatoms due to strain and binding energy related effects, increasing the probability of In desorbing from the surface or diffusing away from the side facets during the remainder of the rotation cycle.

Finally, note that the PL spectra of samples B--E show a slight contribution from the GaN core. The inset in figure~\ref{fig5}(c) compares the GaN core luminescence of the template NWs with that of samples~D and E at 300~K. The peaks of the two (In,Ga)N core-shell samples are clearly red-shifted with respect to the GaN luminescence of the template NWs. Due to the lattice mismatch of GaN and (In,Ga)N, the (In,Ga)N shells impose a tensile strain on the GaN core, which lowers the bandgap energy and shifts the core emission. The strained GaN cores suggest that the shells are generally free of plastic relaxation.

\subsection{Diameter dependence of the NW emission}\label{sec:diam}

\begin{figure}
\centering
\includegraphics[width=0.85\columnwidth, trim= 0 0 0 0]{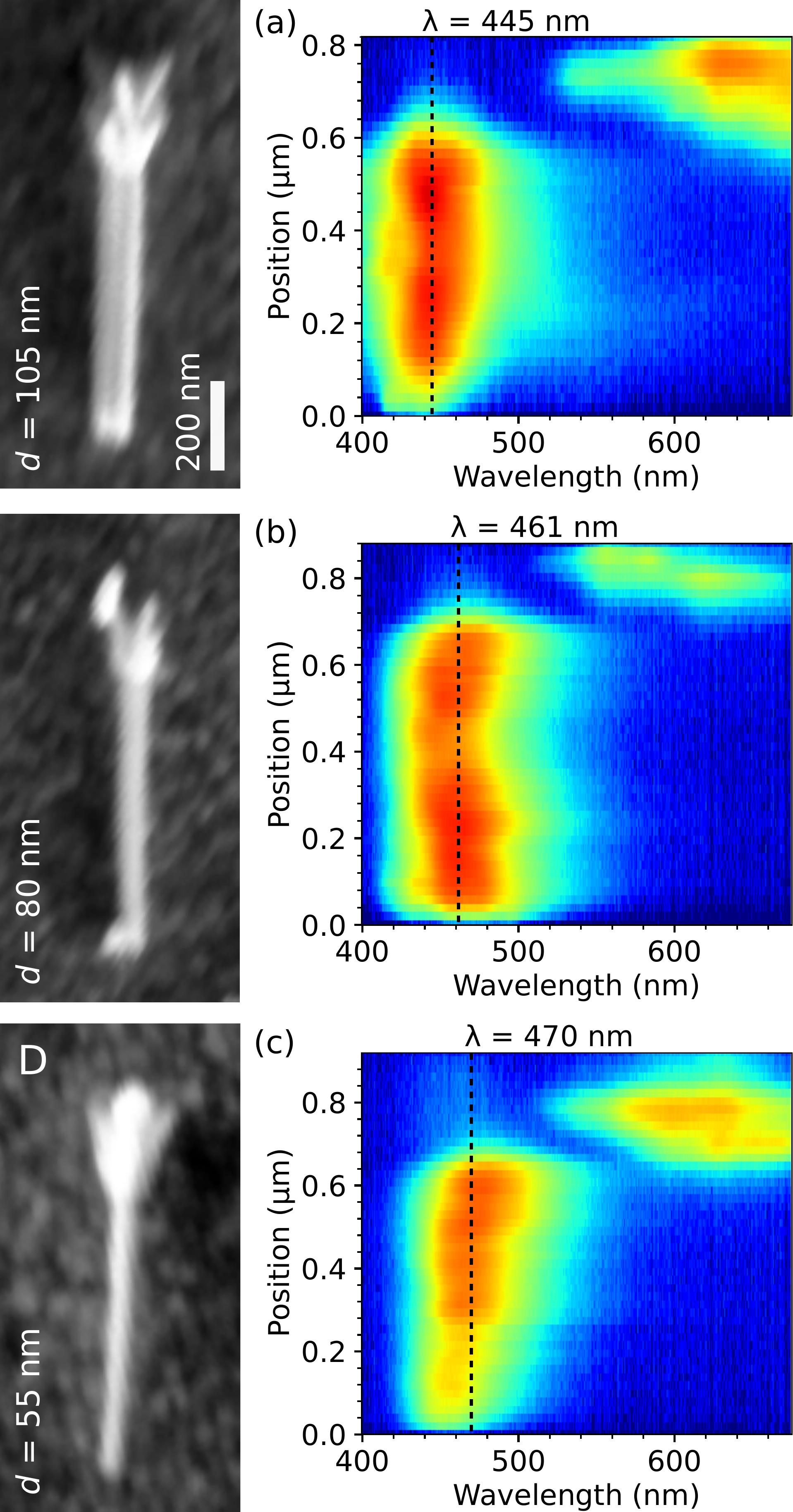}
\caption{Comparison of CL line scans measured at 10~K for NWs of different diameter $d$, as indicated on the respective SE micrographs, for sample D. The average wavelength $\lambda$ of the shell emission, corresponding to the dashed lines, is indicated for each line scan.}\label{fig6}
\end{figure}

Finally, we have looked more closely at the variation in emission between individual NWs from a single sample. The self-assembled growth of the template NWs yields an ensemble with a diameter variation \cite{vanTreeck2018}. Different diameters of the core-shell NWs are thus dominated by changes in the core diameter and not by changes of the shell thickness. Figure~\ref{fig6} compares CL line scans for three different nanowires from sample D with distinctly different diameters, where figure~\ref{fig6}(a) corresponds to the NW already shown in figure~\ref{fig5}(a). The average emission wavelength changes from 445~nm (2.79~eV) for the NW with $d=105$~nm to 470~nm (2.64~eV) for $d=55$~nm. In addition, the latter NW exhibits a slight gradient of the emission wavelength along the NW axis with a redshift towards the NW tip, as already discussed above for sample B. In the ensemble PL spectrum in figure~\ref{fig5}(c), the thickest NW corresponds to the peak wavelength, while the thinner NWs contribute to the broad long wavelength shoulder of this spectrum.

This shift in emission energy by 150~meV with changing NW diameter may be attributed to strain-related effects. First, the (In,Ga)N of the thick NWs is compressively strained from the GaN core. With decreasing core diameter, and thus a more dominant (In,Ga)N volume, this strain will be reduced corresponding to a redshift of the emission. Secondly, the reduced strain will facilitate a higher In incorporation further redshifting the emission of thin NWs.

\section{Conclusions}

The present investigation of (In,Ga)N shells shows that the principal understanding of the core-shell growth by MBE developed in Ref.~\citenum{vanTreeck2020} for GaN shells around GaN NWs can also be transferred to the case of ternary (In,Ga)N shells, providing an understanding of the influence of the various growth conditions on the NW morphology and the In incorporation into the shell. During each rotation round, the side facets pass through different phases---from the wetting phase, through the metal-rich and N-rich growth phases, to the dissociation phase. With In acting as a surfactant, the ternary (In,Ga)N shells are much more homogeneous in thickness along the NW length than pure binary GaN shells. The In content of the (In,Ga)N shells depends on the growth conditions, namely the V/III and the In/Ga ratios, as well as the rotation speed, which determines the duration of the different rotation phases. For a high In incorporation, a high V/III ratio and a fast rotation are beneficial. An investigation of the influence of different azimuthal angles between the different material sources on the shell growth was beyond the scope of this work, but may be interesting to further optimise the shell growth.

In general, we have demonstrated that the growth by MBE of (In,Ga)N shells emitting in the green spectral range for NWs with a final diameter below 100~nm is feasible. Furthermore, our work deepens the understanding of the processes underlying (In,Ga)N shell growth by MBE and stresses the importance of the geometrical source arrangement. As an outlook, active regions emitting in the red could be within reach. Remaining challenges in using such core-shell structures as the basis for NW LEDs, are the length variation of the template NWs and the large top segments, both of which may adversely affect the processing of the final LED device.

\ack
The authors would like to thank Neha Aggarwal for a critical reading of the manuscript and Sergio Fernández-Garrido for fruitful discussions. Furthermore, the authors are grateful to Katrin Morgenroth, Carsten Stemmler and Michael Höricke for the maintenance of the MBE system, as well as to Anne-Kathrin Bluhm for the SE micrographs.

\bibliography{vanTreeck_nt}

\end{document}